\documentclass[pre,twocolumn,a4paper,showpacs]{revtex4}
\usepackage{amssymb}
\usepackage{epsfig}
\begin{document}
\title{Mass media destabilizes the cultural homogeneous regime in
Axelrod's model}

\author{Lucas R. Peres and Jos\'e F. Fontanari}
\affiliation{Instituto de F\'{\i}sica de S\~ao Carlos,
  Universidade de S\~ao Paulo,
  Caixa Postal 369, 13560-970 S\~ao Carlos, S\~ao Paulo, Brazil}

\begin{abstract}
An important feature of Axelrod's model for culture dissemination or social influence is the emergence of many
multicultural absorbing states, despite the fact that the local rules that specify the agents interactions are
explicitly designed to decrease the cultural differences between agents. Here we re-examine the problem of introducing 
an external, global interaction -- the
mass media  -- in the  rules of Axelrod's model: in addition to their nearest-neighbors, each agent
has a certain probability $p$ to interact with a virtual neighbor whose cultural features are fixed from the outset.
Most surprisingly, this apparently homogenizing effect actually increases
the cultural diversity of the population. We show that, contrary to previous claims in the literature, even a
vanishingly small value of $p$ is sufficient to destabilize the homogeneous regime for very large
lattice sizes. 
\end{abstract}

\maketitle

%
\section{Introduction} \label{sec:Intro}

Why do people have different opinions given that after repeated interactions 
some consensus should emerge?  Why are there different cultures given
that modern media has apparently succeeded in transforming the planet into a global village \cite{McLuhan_66}? 
These are the issues addressed by  Axelrod's model 
for the dissemination of culture or social influence \cite{Axelrod_97}, which is considered the  
paradigm for idealized models of
collective behavior which seek to reduce a collective phenomenon to its functional essence \cite{Goldstone_05}. 
In fact, building on just a few simple principles, Axelrod's model  provides 
highly nontrivial answers to those questions. In Axelrod's model, an agent 
-- an  individual or a culturally homogeneous village --
is represented by a string of $F$ cultural features, where each feature can adopt a certain number $q$ of distinct traits. 
The interaction between any two agents takes place
with probability proportional to their cultural similarity, i.e., proportional to the number of 
traits they have in common. The result of such interaction 
is the increase of the similarity between 
the two agents, as one of them modifies  a previously  distinct trait to match that of its partner.

Notwithstanding the built-in assumption that social actors have a tendency to become more similar 
to each other through
local interactions \cite{Latane_81,Moscovici_85}, Axelrod's model does exhibit global polarization, i.e., a stable 
multicultural regime \cite{Axelrod_97}. More importantly, however, at least from the statistical physics perspective, is the fact that
the competition between the disorder of the initial configuration and  the ordering bias of the local  interactions produces
a nontrivial threshold phenomenon (more precisely, a nonequilibrium phase transition) which separates  in the space of 
parameters of the model the globally homogeneous from
the globally polarized regimes \cite{Castellano_00,Lauro}. 

A feature that sets  Axelrod's model apart from most
lattice models which exhibit nonequilibrium phase transitions \cite{Marro,review} is the fact that all stationary 
states of the dynamics are absorbing states, i.e., the dynamics freezes in the long time regime \cite{Castellano_00}.
This is so because, according to the rules of Axelrod's model, two neighboring agents who do not have any cultural trait 
in common cannot interact and the interaction between agents who share all the cultural traits does not change
their cultural features. Hence at equilibrium we can safely predict that, regarding their cultural features,
any neighbor of a given agent is either identical to or completely different from it.
This is a double-edged sword:  on the one hand, we can easily identify the stationary regime, which is a major
problem in the characterization of nonequilibrium phase transitions; on the other hand,  the
dynamics can take an arbitrarily large time to freeze for some
parameter settings and initial conditions \cite{Castellano_00,Lauro,Vilone_02,Vazquez_07}.

The key ingredient for the existence of a stable  globally polarized 
state  is  the rule that prohibits the interaction between completely different agents (i.e., agents which
do not have a single cultural trait in common).
This was first pointed out by Kennedy \cite{Kennedy_98} who relaxed this rule and permitted interactions
regardless of the similarity between agents. As a result, the system evolved until
all agents became identical, i.e., the only absorbing states were the homogenous ones. (There are $q^F$ distinct
absorbing homogenous configurations.) In addition, Klemm et al. \cite{Klemm_03c} 
have shown that the introduction of external noise to the dynamics so that a single trait of an arbitrarily  chosen agent 
was changed at random ends up destabilizing the polarized state. Moreover, expansion of communication modeled by 
increasing the connectivity of the lattice \cite{Greig_02,Klemm_03b} or by  placing the agents in more complex networks \cite{Klemm_03a}
(e.g., small-world and scale-free networks) also resulted in cultural homogenization.

It should be mentioned, however, that other models of social influence seem to yield a more robust polarized state.
For instance, the frequency bias mechanism \cite{Boyd_85,Nowak_90}  for cultural or opinion change assumes that 
the number of people holding an opinion is the key factor for an agent to adopt that opinion, i.e., people
have a tendency to espouse cultural traits that are more common in their social environment. This is
then the standard voter model of statistical physics \cite{Ligget}. Parisi et al. \cite{Parisi_03} 
have replaced the rules of Axelrod's model by the frequency bias mechanism (essentially, a majority rule)
and found a stable polarized state for small lattices. Since similarity plays no role in the
agents' interactions, the frequency bias mechanism is naturally robust to noise.

The impression is then that the globally polarized (multicultural) state is very frail, being disrupted by
any (realistic or not) extension of the original model. In view of this feeling, it came as a big surprise
the finding by  Shibanai et al. \cite{Shibanai_01}  that the introduction  of a homogeneous media effect
(i.e., it is the same for all agents) aiming at  influencing the agents' opinions actually favors polarization.
This finding is at odds with the common-sense view that mass media, such as newspapers and television, are devices 
that can be effectively used to control people's opinions and so homogenize society. Of course, the effect of media in
real personal networks is complicated and seems to follow the so-called `two-step flow of communication' in which 
the media affect opinion leaders first, who then influence the rest of the population \cite{Lazarsfeld}. In fact, personal
networks seem to serve as a buffer for the media effect.

Although this counterintuitive effect of the mass media has been extensively investigated
(see, e.g., \cite{Avella_05,Avella_06,Mazzitello_07,Candia_08}) there is still no first-principles explanation
to it. The research has focused mostly on the search for a threshold on the intensity of the media influence such
that above that threshold the population would become polarized and below it, the population would becomes culturally homogeneous.
In this contribution we show that such threshold is in fact an artifact of finite lattices: when
a careful analysis of the finite-size effects  is carried out we find that even a vanishingly small media influence  is
sufficient to destabilize the culturally homogeneous regime.

The rest of this paper is organized as follows. In Sect.~\ref{sec:model} we describe the original Axelrod's model,
discuss at some length the basic assumptions of the model and introduce the effect of an external fixed media 
\cite{Avella_05,Avella_06}. In Sect.~ \ref{sec:res} we present an efficient 
algorithm to simulate Axelrod's model. The simulation results as well as a discussion of our main results are
presented also in that section. Finally, in Sect.~ \ref{sec:conc} we present our concluding remarks.

\section{Model}\label{sec:model}

In Axelrod's model each agent is characterized by a set of $F$ cultural features which can take on
$q$ distinct values. Hence an agent is represented by a string of symbols, e.g. $13255$  in the
case of $F=5$ and $q=5$. Clearly, for this parameter setting there are only $q^F = 3125$ different cultures.
The agents are fixed in the sites of a square lattice of size $L \times L$ 
with periodic boundary conditions  and can interact only
with their four nearest neighbors.
The initial configuration is completely random with the features of each agent given by  
random integers drawn uniformly from $1$ to $q$. At each time we pick an agent at random
(this is the target agent) as well as one of its neighbors. These two 
agents interact with probability equal to  their cultural similarity, defined as the fraction of 
common cultural features. For instance, assuming that the target agent is described by the string
$13255$ and its neighbor by $13425$, the interaction occurs with probability $3/5$. In case the interaction
action is not selected, we pick another target agent at random and repeat the procedure.
An interaction consists of selecting at random one of the distinct features, and changing the
target agent's trait on this feature to the neighbor's corresponding trait. Returning to our example,
if the third feature is chosen the target agent becomes $13455$ and its neighbor remains unchanged.
This procedure is repeated until the system is frozen in an absorbing configuration. 

The basic assumption of Axelrod's model is that similarity is a main requisite for social interaction
and, as a result, exchange of opinions. This is the `birds of a feather flock together' hypothesis which
states that individuals who are similar to each other are more likely to interact
and then become even more similar \cite{Moscovici_85}.  Recent empirical evidence in favor of this assumption comes
from  the analysis of Web 2.0 social networks \cite{Singla_08}. Study of a population of over $10^7$ people indicates 
that people who chat with each other using instant messaging are more likely to have common interests,
as measured by the similarity of  their Web searches, and the more time they spend talking,
the stronger this relationship is. We note, however, that this assumption is disputed by other researchers  who
argue that people are attracted to others who resemble their ideal, rather than their actual selves \cite{Wetzel_82}.

To introduce the effect of a global media following the seminal paper by Shibanai et al. \cite{Shibanai_01}, we need first
to define a virtual  agent whose cultural traits reflect the media message. In Ref.\ \cite{Shibanai_01}, each 
cultural feature of the virtual agent has the trait which is the most numerous in the population --
the consensus opinion. Here we choose to keep the media message
fixed from the outset, so it really models some alien influence impinging on the population.
Explicitly, we generate the culture vector of the virtual agent at random and keep it fixed
during the dynamics.
Next, we need to specify how the media interact with the real agents.  To do that we introduce a new control parameter
$p \in \left [ 0,1 \right ] $, which measures the strength of the media influence. As in the original Axelrod's
model, we begin by choosing  a target agent at random, but now it  can interact with the media with probability $p$ 
or with its neighbors with probability $1-p$.  Since we have defined the media as a virtual  agent, 
the interaction follows exactly the same rules as before. The original model is recovered for  $p=0$, provided
we properly define  the halting criterion of the dynamics, as  discussed in the next section.

\section{Results}\label{sec:res}

To simulate efficiently Axelrod's model we  make a list of the active agents. An active agent is an agent 
that has at least one feature in common  and at least one feature distinct with at least one of its four nearest neighbors. 
Clearly, since only active agents can change their cultural features, it is more efficient  to select the
target agent randomly from the list of active agents rather than from the entire lattice. Note that the
randomly selected neighbor of the target agent may not necessarily be an active agent itself. In the case that
the cultural features of the target agent are modified by the interaction with its neighbor, we 
need to re-examine the active/inactive status of the target agent as well as of all its neighbors so as
to update the list of active agents.  The dynamics is frozen when the list of active agents is empty. This is
the halting criterion we mentioned in the last section. 

The important point in this halting criterion is that the virtual agent does not enter
the procedure to determine  whether a real agent is active or not; otherwise the dynamics would not freeze.
Actually, there are only two situations where the dynamics could freeze in the case the virtual
agent is used in that procedure: in the uniform regime where all agents become identical to the virtual agent,
and in a two-domains regime where one domain is identical to the virtual agent and the other is completely opposed (there are
$\left ( q-1 \right )^F$ distinct realizations of this possibility). However, since the dynamics does not lead in
general to these situations, it gets stuck in a trite position in which changes occurs  due to the interaction with
the virtual agent only. Although it has never been explicitly pointed out, this must have been the halting criterion  used in previous analyses
of the effect of media in Axelrod's model \cite{Shibanai_01,Avella_05,Avella_06,Mazzitello_07,Candia_08}.
 
In  order to
explore fully the dependence of the frozen configuration on the lattice size, in this contribution we restrict our analysis 
to the parameter setting $F=q=5$. 
A feature that sets our results apart from those reported previously in the literature is that
our data points represent averages over at least $10^3$ independent runs for lattices of
linear size up to $L=3000$. (For comparison, we note that the results of Refs.\ \cite{Avella_05,Avella_06} are
derived from simulations of lattices with $L=40$ and $50$ independent runs.)  This requires a substantial
computational effort, especially in the regime where the number of cultures decreases with
the lattice size since then the time for absorption can be as large as $10^6 \times L^2$.  
In the figures presented in the following,
the error bars are smaller or at most equal to the symbol sizes.

\begin{figure}
\centerline{\epsfig{width=0.52\textwidth,file=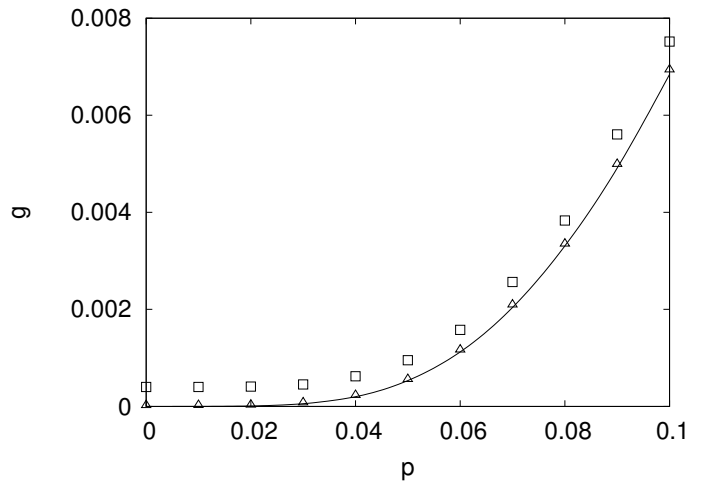}}
\par  
\caption{Ratio $g$  between the number of cultural domains and the lattice area  as function of the
strength of the media influence  for  $L= 50 \left( \square \right)$ and 
$200 \left( \triangle \right)$. The solid line is the result of
the extrapolation of the data to the limit $L^2 \to \infty$. The parameters are  $F=5$ and $q=5$.
\label{fig:1} }
\end{figure}

For our purposes, the frozen configuration can be characterized by the 
ratio between the number of clusters  (or cultural domains) $S$ and the lattice area $L^2$. A cluster is simply
a bounded region of uniform culture. In the case of diasporas \cite{Greig_02}, the two or more cultural domains (which are
characterized by the same culture) are counted separately. We note that since $S$ is bounded by $L^2$ we have  
$g \equiv S/L^2 \leq 1$. In the uniform regime we have $S=1$ and so $g = 1/L^2$. Figure \ref{fig:1} exhibits this 
measure as function of the strength of the media influence $p$ for different lattice sizes. 
The suitability of the measure $g$  is demonstrated by the fact that the data 
converge to well-defined values (solid line in Fig.\ \ref{fig:1}) as the lattice size is increased. In other
words, $S$ increases with $L^2$ for $p$ not  too small. Indeed, from Fig.\ \ref{fig:1} it seems that the measure $g$ vanishes
for small $p$ which would indicate the existence of a minimum strength value $p_c$, above which the
uniform regime is destabilized \cite{Avella_05,Avella_06,Mazzitello_07,Candia_08}. Visual inspection of the data
shown in Fig.\ \ref{fig:1} yields  $p_c \approx 0.03$, which agrees with the estimate of Ref.\ \cite{Avella_06}
(see their Figure 3).

\begin{figure}
\centerline{\epsfig{width=0.52\textwidth,file=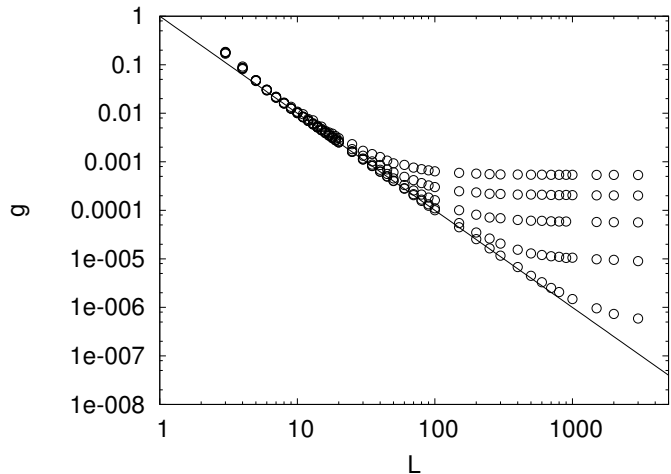}}
\par
\caption{Logarithmic plot of the ratio $g$ as function of the linear
size $L$ of the lattice for (top to bottom) $p=0.05,0.04,0.03,0.02$ and $0.01$. The 
solid straight line is $1/L^2$, which corresponds to the value of $g$ in the uniform regime.
The parameters are  $F=5$ and $q=5$.
\label{fig:2} }
\end{figure}

A more careful analysis reveals a different story, however, as shown in Fig.\ \ref{fig:2}. In fact, 
consider the data for $p=0.01$, which is well below our initial estimate, $p_c \approx 0.03$.
An analysis of lattices of sizes up to $L=600$ indicates a clear tendency of convergence
towards the uniform regime (i.e., $g=1/L^2$ fits the data almost perfectly in that range of $L$), but this trend changes completely 
when lattices of sizes greater than $L=1000$ are considered. In this case, rather than vanishing as $1/L^2$,  $g$  tends to a nonzero value 
when $L \to \infty$. To verify whether this finding holds true for all
values of $p$  we need first estimate the value of
$g=g \left ( p \right) $ for infinite lattices and nonzero $p$ and then  try to figure out the  
dependence of the extrapolated value of $g$ on $p$ in the limit  $p \to 0$.  As suggested by Fig.\ \ref{fig:2}, direct simulations 
using small values of the parameter $p$  would require very large lattice sizes in order to produce
significative  deviations from the uniform regime.

\begin{figure}
\centerline{\epsfig{width=0.52\textwidth,file=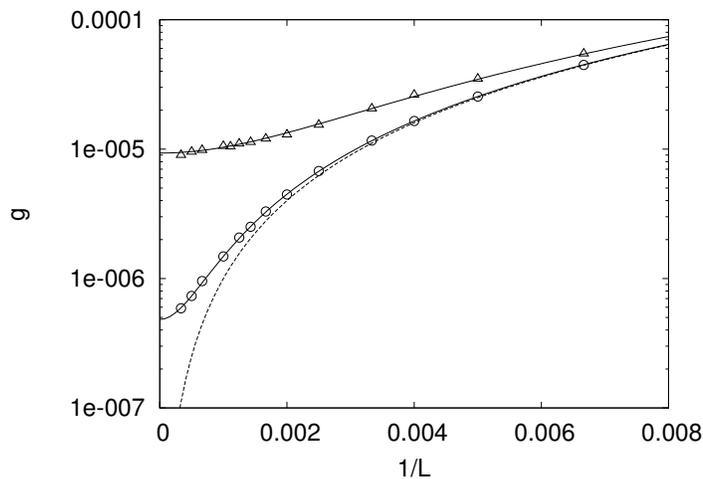}}
\par 
\caption{Ratio $g$  between the number of cultural domains and the lattice area as function of the
reciprocal of the lattice size $1/L$ for $p=0.02 \left ( \triangle \right )$ and $0.01 \left ( \circ
\right )$. The solid lines are the fittings $g = B_p + A_p/L^2$ and the dashed line is the function
$1/L^2$. The parameters are  $F=5$ and $q=5$.
\label{fig:3} }
\end{figure}

\begin{figure}
\centerline{\epsfig{width=0.52\textwidth,file=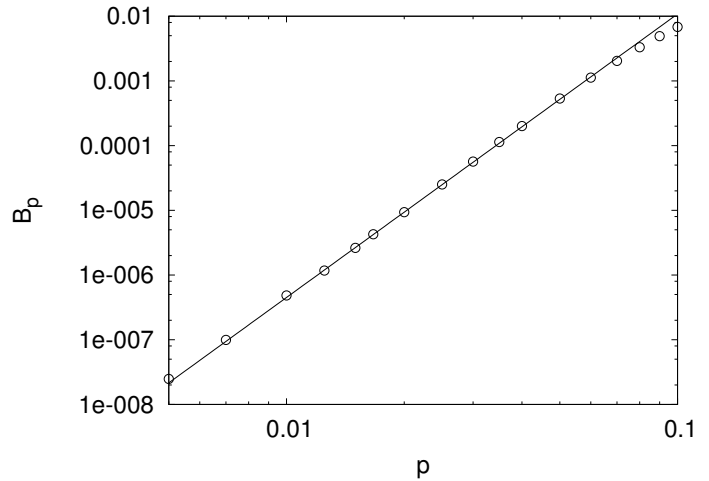}}
\par 
\caption{The ratio between the number of cultural domains and the lattice area for $L \to \infty$ 
obtained through the extrapolation procedure shown in Fig.\ \ref{fig:3}
as function of the strength $p$ of the media influence.
The straight line is the fitting given by Eq.\ (\ref{fit}). The parameters are  $F=5$ and $q=5$.
\label{fig:4} }
\end{figure}

Figure \ref{fig:3} illustrates the procedure used to obtain the measure $g$ in the limit $L \to \infty$. The key point 
is the use of the fitting  function $g \left ( p \right )  = B_p + A_p/L^2$ which describes the data very well 
for $L > 500$: the statistical error in the estimate of $B_p = \lim_{L \to \infty} g \left ( p \right ) $ 
is less than $2\%$ for all values of $p$ considered here. The solid curve shown in Fig.\ \ref{fig:1} was obtained
following this procedure. Finally, Fig. \ref{fig:4} presents the dependence of $B_p$ on $p$. 
For small $p$  the data is  fitted very well by the equation 
\begin{equation}\label{fit}
B_p = \lim_{L \to \infty} g \left ( p \right ) = \left ( 260 \pm 29 \right ) p^{4.38 \pm 0.03}
\end{equation}
as indicated in the figure. The large value of the power of $p$  may explain why the numerical 
simulations yielded a nonzero value
for $p_c$: for small $p$ it is virtually impossible to distinguish the result of Eq.\ (\ref{fit}) from
zero. 

In sum, Axelrod's model does not exhibit a phase transition for $p > 0$: the only 
stable regime for infinite lattice
sizes is the polarized one.  Strictly, this conclusion is valid for a single
setting of the control parameters, namely, $F=q=5$, but we see no reason why it should
not hold for other values of these parameters as well.

\section{Conclusion}\label{sec:conc}

In this contribution we have revisited an important extension of Axelrod's model in which, in addition to
the local interactions between agents, there is a global element  -- the media --  that influences the agents'
opinions or cultural traits \cite{Shibanai_01}. In stark contrast to the common sense opinion that the media effect is to homogenize 
the society we find, in agreement with previous studies \cite{Shibanai_01,Avella_05,Avella_06,Mazzitello_07,Candia_08},
that the media actually promotes polarization or the diversity of opinions. However, we have shown that this effect is
so powerful that a vanishingly small influence strength $p$ is sufficient to destabilize the cultural homogenous state for
very large lattices. This finding calls for a re-examination of the claim, which is based on the analysis of small
lattices, that there exists a threshold value $p_c$ below which the homogeneous state is stable.
At present we have no idea of why the media promotes polarization rather than the expected homogenization. 
An analysis of
the distribution of sizes of the cultural domains as well as of the distance between domains may provide some cue on this counterintuitive effect. 
Work in this line is under way.

\begin{acknowledgments}
This research  was supported by CNPq and FAPESP, Project No. 04/06156-3.
\end{acknowledgments}


\end{document}